\documentclass[pre,showpacs,twocolumn,superscriptaddress]{revtex4}
\usepackage{amssymb,graphicx}
\begin{document}

\title{Flow improvement caused by agents who ignore traffic rules}
\author{Seung Ki Baek}
\affiliation{Department of Physics, Ume{\aa} University, 901 87 Ume{\aa},
Sweden}
\author{Petter Minnhagen}
\affiliation{Department of Physics, Ume{\aa} University, 901 87 Ume{\aa},
Sweden}
\author{Sebastian Bernhardsson}
\affiliation{Department of Physics, Ume{\aa} University, 901 87 Ume{\aa},
Sweden}
\author{Kweon Choi}
\affiliation{Gyeonggi Science High School, Suwon 440-800, Korea}
\author{Beom Jun Kim}
\email[Corresponding author, E-mail: ]{beomjun@skku.edu}
\affiliation{ Department of Energy Science and BK21 Physics Research Division,
Sungkyunkwan University, Suwon 440-746, Korea}
\affiliation{Department of Computational Biology, School of Computer Science
and Communication, Royal Institute of Technology, 100 44 Stockholm, Sweden}

\begin{abstract}
A system of agents moving along a road in both directions is studied
numerically within a cellular-automata formulation.
An agent steps to the
right with probability $q$ or to the left with $1-q$ when encountering
other agents.
Our model is restricted to two agent types, traffic-rule abiders ($q=1$)
and traffic-rule ignorers ($q=1/2$). The traffic flow, resulting from
the interaction between these two types of agents, is obtained as a
function of density and relative fraction.
The risk for jamming at a fixed
density, when starting from a disordered situation,
is smaller when every agent abides by a traffic rule than when all
agents ignore the rule. Nevertheless, the absolute minimum occurs when
a small
fraction of ignorers are present within a majority of abiders. The
characteristic features for the spatial structure of the flow pattern are
obtained and discussed.
\end{abstract}

\pacs{89.40.--a,45.70.Vn,87.23.Ge,02.50.Le}


\maketitle

\section{Introduction}

Society has various rules to regulate interactions among its
members. While some of these rules might be enforced by
authorities and turned into laws, quite a few have evolved over a long time
and have turned into conventions.
Just as in natural sciences, rules or conventions, regulating the
interaction between individual constituents, often result in emerging global
patterns.
A traffic rule which enforces individual vehicles/pedestrians to move along
only one side of
a road clearly results in a global traffic flow pattern (see, e.g.,
Ref.~\cite{young}).
Because of this connection, traffic problems have often been studied by
using methods and concepts from
nonequilibrium statistical physics~(for a review, see
Refs.~\cite{chow} and \cite{helbing}).
The approaches from physics include hydrodynamic
descriptions~\cite{hydro}, differential equations describing effective
microscopic forces~\cite{ode}, and cellular
automata (CA)~\cite{biham,fukui,mura,automata,lee2004a}.
In particular, the CA approach is often used in broad
contexts of agent-based modeling as an efficient way of accounting for
complicated interactions among constituents. Due to the computational
efficiency, CA is particularly suitable for analyzing
the dynamics of many individuals who try to move in different directions
while at the same time being influenced by 
the motions of other individuals.
It is notable that the jamming in vehicular traffics has natures different
from that in pedestrian traffics. The former is explained by the time delay in
the responses of the drivers, and this is the reason why the jamming may easily
occur with vehicles only in one direction~\cite{lee2004a}.
In the latter case, on the other hand, the jamming is caused by the
collision of agents in opposite directions~\cite{fukui}. This study is
mainly focused on this pedestrian case.

In the present work, we use the CA approach and find that the minimal risk for
a jamming of the pedestrian flow occurs when a small fraction of traffic-rule
ignorers is present within a majority of traffic-rule abiders. Even though
this result is obtained within our simplified model system, it raises an
interesting question on the observability and implication of such a phenomenon
in social systems. Here we provide a detailed description on this
observation as well as a qualitative understanding.

This paper is organized as follows. In Sec.~\ref{sec:coord}, we
review the basics of a coordination game based on a traffic rule.
We describe how we have performed our numerical
experiments in
Sec.~\ref{sec:numset}. The results are presented and compared to the
coordination game in Sec.~\ref{sec:res}. The results are summarized in
Sec.~\ref{sec:dis}.

\section{Coordination game}
\label{sec:coord}

Let us first consider two players moving on a road in opposite
directions heading for a direct collision. Each of them can
choose to step aside left ($L$) or right ($R$) in order to avoid the collision
and it is avoided only if both make the same choice. 
Thus the options both $L$ or both $R$ are equally gainful, whereas the
choices $LR$ and $RL$ lead to collision.
The situation is summarized
in Table~\ref{table:coord} in the form of a
doubly-symmetric two-person coordination game~\cite{davis,weibull}. 
Each player will behave according to a strategy in the form of a
complete description of which action is taken under every possible circumstance.
In what follows, we denote the strategy of an agent as $S_q$ if she
chooses $R$ with probability $q$ (and thus $L$ with probability $1-q$).
For example, if a player always chooses $R$ ($L$), her strategy is
represented as $S_1$ ($S_0$).

\begin{table}
\caption{Normal form of a traffic-rule coordination game. Since the situation
is symmetric to each player, only one player's payoff is presented in this
table after normalized to unity. Both players are better off when they
choose the same moves (left/left and right/right).}
\begin{tabular*}{\hsize}{@{\extracolsep{\fill}}ccc}\hline\hline
 & Left & Right\\\hline
Left & 1 & 0 \\
Right & 0  & 1 \\\hline\hline
\end{tabular*}
\label{table:coord}
\end{table}

The concept of equilibrium is useful in analyzing a game: suppose that everyone
has chosen a strategy so that no one gains anything by changing her
strategy unilaterally, such a set of strategies constitute a {\em Nash
equilibrium}~\cite{weibull}. In the case of the coordination game, there exist
three strategies which are Nash equilibria, i.e., $S_1$, $S_0$, and $S_{1/2}$.
The pure strategies $S_1$ and $S_0$ simply represent the ordinary 
traffic rules such that agents should always step aside to the
right or always to the left.
Due to the left-right symmetry in the problem both $S_1$ and $S_0$ 
compose Nash equilibria. On the other hand, if one makes a
decision at random by tossing a coin, then obviously the opposing player cannot 
gain anything no matter what strategy she changes to. Consequently, 
the mixed strategy $S_{1/2}$ also constitutes a Nash equilibrium.

We next consider the evolutionary stabilities~\cite{maynard} of these Nash
equilibria in a population
where every pair of members plays the game.
Suppose that almost all the players adopt a certain strategy $S$. The
strategy $S$ is called evolutionarily stable when another mutant
strategy $\tilde{S}$ cannot invade the population of $S$ 
since the payoff of $\tilde{S}$ is less than that of $S$. Mathematically,
the evolutionary stability of the mixed strategy $S_q$ is equivalent to
the stability of a population where a fraction $Q=q$ of members have $S_1$
while the others have $S_0$~\cite{maynard}. Note that such equivalence holds
only when there exist two pure strategies.
In the stability analysis, one often employs a dynamics resulting as
individuals in a group try to adopt the strategies of more successful
individuals. Such a situation can be modeled as follows:
the relative proportion $Q_S$ of players who use strategy $S$ is assumed
to 
grow in time in proportion to the payoff $W_S$ at the last time step. This
particular dynamics is given by
the {\em replicator dynamics} equations~\cite{maynard},
\[
\dot{Q_S} \equiv \frac{dQ_S}{dt} = Q_S\left(W_S-\sum_{S'} W_{S'} Q_{S'}\right)
\]
within the continuum time approximation, where the last term has been
inserted to make the constraint $\sum_S Q_S = 1$ fulfilled at any time $t$.
In our traffic-rule game, we have
$Q_{S_0} + Q_{S_1} = 1$, and therefore we may set $Q_{S_1} \equiv Q$ and
$Q_{S_0} \equiv 1-Q$ to study the evolutionary stability of $S_q$ with
$q=Q$. From Table~\ref{table:coord}, it follows that the expected
pay-off for an agent is given by the probability of encountering a
traffic-rule abider or ignorer, respectively, resulting in
$W_{S_1}=Q$ and $W_{S_0}=1-Q$. For example, if $Q=1/4$, an agent with
strategy $S_1$ will have a chance out of four to meet another with the same
strategy, meaning that her expected payoff amounts to $1/4$ at every
encounter.
Consequently, the replicator dynamics 
equations can be cast in the form
\begin{eqnarray*}
\dot{Q} &=& Q \{Q - [Q^2 + (1-Q)^2]\}\\
        &=& Q (1-Q) (2Q-1).
\end{eqnarray*}
From the stationarity condition $\dot{Q} = 0$ 
we find three Nash equilibria at $Q_1=0$, $Q_2 = 1$, and
$Q_3 = 1/2$, which in turn correspond to $q=0, 1$, and $1/2$, respectively.
The linear perturbation $\epsilon_n$ introduced to the $n$th Nash equilibrium 
by $Q = Q_n + \epsilon_n$ ($n=1,2,3$) satisfies 
$\dot{\epsilon_1}/\epsilon_1 < 0$, 
$\dot{\epsilon_2}/\epsilon_2 < 0$,  and
$\dot{\epsilon_3}/\epsilon_3 > 0$, respectively, and we find that only
the last equilibrium point $Q=1/2$ is {\em unstable} in this dynamics.
In other words, a convention of randomly choosing left or right is
unlikely to emerge, since
people eventually learn that the traffic improves if a majority settles for
one of the alternatives.
Note that this conclusion is based on the assumption of full mixing,
corresponding to the mean-field approximation in physics.
To what extent is the simple picture,
implied by Table~\ref{table:coord}, also valid for a two-dimensional
plane filled with moving agents? This question is investigated in the
following.

\begin{figure}
\includegraphics[width=0.15\textwidth]{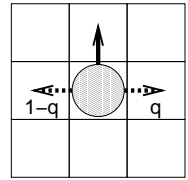}
\caption{Movements of a self-propelled agent.
The solid arrow
indicates the agent's intrinsic direction, while the dotted ones indicate its
possible evasions when the intrinsic direction is blocked by another
agent.}
\label{fig:model}
\end{figure}

\section{Numerical setup}
\label{sec:numset}

\subsection{Moving code}

We start with a simple self-propelled agent,
derived from Ref.~\cite{mura}, which obeys the following moving code (compare
to Fig.~\ref{fig:model}). It has its own intrinsic direction, either
upward or downward. It can move only to a neighboring cell at each time
step, and a single cell
cannot allow  more than one agent at the same time.
If the agent's front cell is empty, it moves to the cell with the
probability $1-s$, where
$s$ denotes the probability of spontaneous stopping.
If the agent is prevented from moving forward, because the front cell is already
occupied, then it steps aside to the right with a probability $q$
or to the left with $1-q$.
In case it attempts to the right (left) but cannot because there is
already an agent in that cell, then it
proceeds to try the alternative option left (right). In case this is also
prevented, it just remains in its present cell.
At least two steady states can be found within this moving code:
one is the
complete jamming where no one can move forward and
the other one is a perfectly
collisionless flow where every
column is occupied only by agents moving in the same direction. 

\subsection{Initialization}

A road is a two-dimensional plane, which has a size of $X \times Y$ in
units of cells.
We impose a periodic boundary (PB) in the $y$ direction to make the road
homogeneous in that direction.
In the $x$ direction, on the other hand, there are walls which prevent
agents from being at $x \leq 0$ or $x > X$.
At the initial time, the agents are randomly distributed with a density
$\rho$ on the road, and their intrinsic directions are given upward or
downward with equal probability.
Such a starting condition is qualitatively similar to a walking street
filled with mingled pedestrians which all start to walk home at the same
moment.
The number of agents is $N=\rho XY$, and the numbers
of upward and downward agents are $N/2$. Among these $N$
agents, $pN$ agents have $q=1$ so that they always try the
right-hand side first, while the others have no preferences, i.e.,
$q=1/2$. There are no initial correlations among the position, intrinsic
direction, and preference.

\subsection{Recursive update}
\label{sub:recur}

All agents make moves in accordance with the 
moving code in a random sequential order (RSO). A simultaneous update is, in
practice, not possible since each update then involves finding all consistent
possibilities based on all individual possibilities of all the agents.  
The RSO update together with the PB condition causes an artifact
called
{\it deadlock}. Imagine that one column is fully occupied by players having
the same intrinsic direction with zero stopping probability ($s=0$).
Even though all of them want to move in the same direction, they
cannot within the RSO update since no one finds an empty space in front of
herself.
Therefore, RSO needs to be modified as follows.
Suppose that an agent $A$ is picked up by RSO to be updated. Then
we regard $A$'s current position as empty and search for a new position for
it according to the moving code.
If agent $A$ is blocked from going forward by another agent $B$,
which has the same intrinsic direction as $A$ but not updated yet at this
time step, we do not exclude the possibility for both of them to move
together simultaneously, so we let $B$ move first.
If $B$ is also in the same situation by a third agent $C$,
this procedure is repeated recursively. When this recursion goes all the
way around PB to $A$'s position again, the column of agents will be
updated by one cell forward altogether.
One time step is completed when the moving code is
applied to all the $N$ agents.

\section{Results}
\label{sec:res}

\begin{figure}
\includegraphics[width=0.45\textwidth]{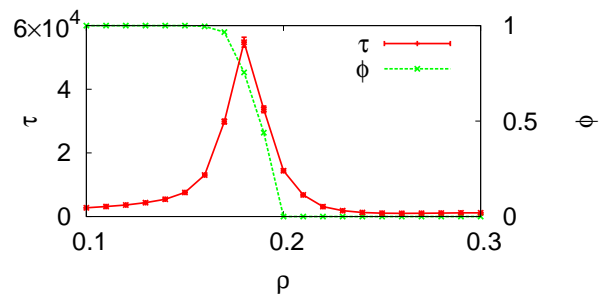}
\caption{(Color online)
Transition between two steady states, the complete jamming and
the collisionless flow, as density $\rho$ changes.
Both of the fraction $p$ of rule abiders and the stopping probability $s$
are set to zero and the road is given as
$50 \times 200$.
Depicted here are the flow $\phi$ and the time $\tau$
to either of the steady states after being averaged over $10^3$
samples under a cutoff of $t_c=10^6$.
}
\label{fig:tran}
\end{figure}

\begin{figure}
\includegraphics[width=0.45\textwidth]{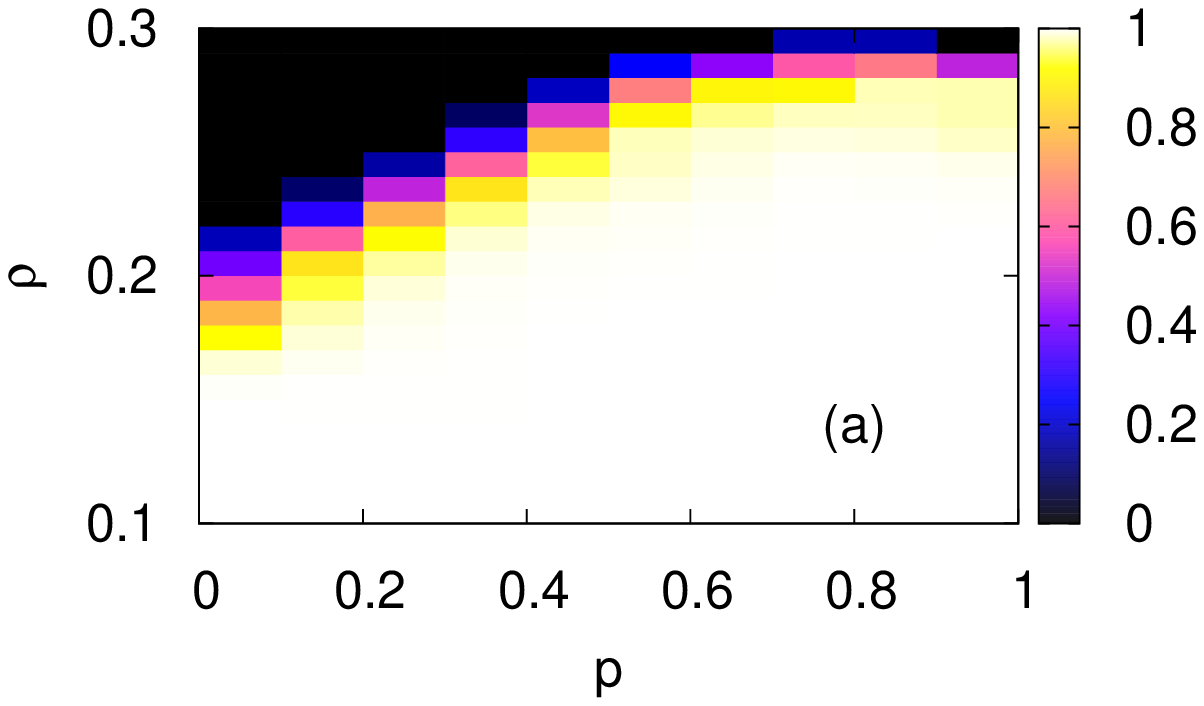}
\includegraphics[width=0.45\textwidth]{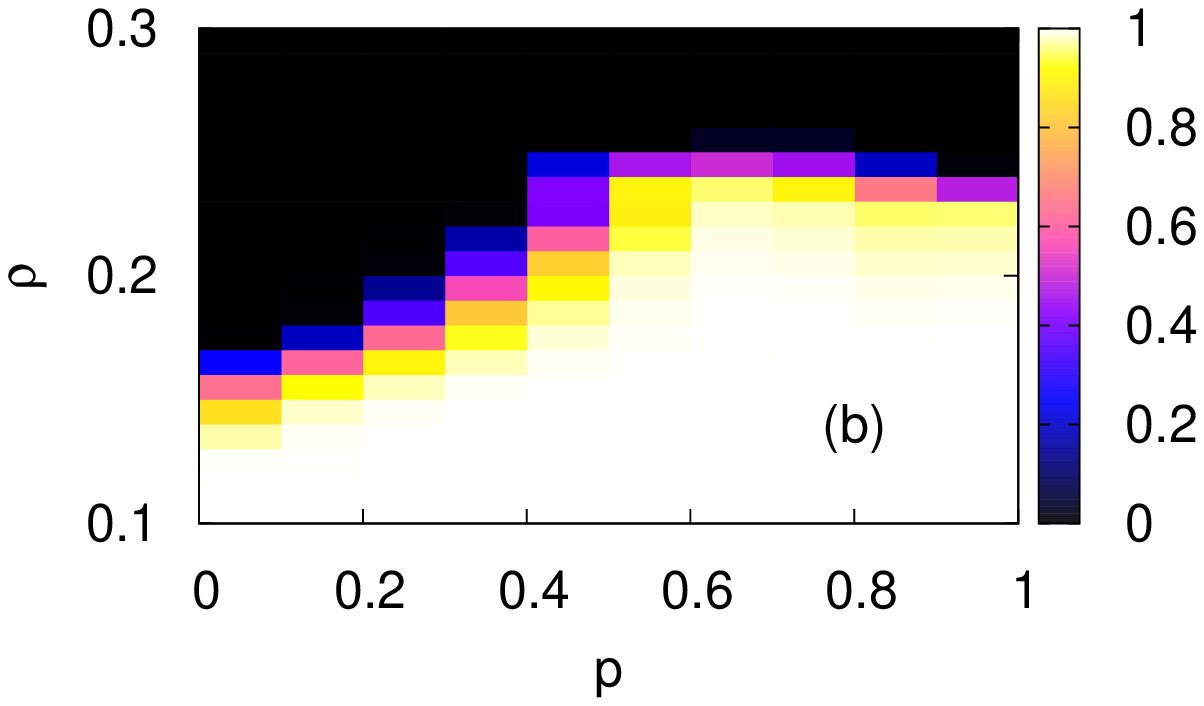}
\includegraphics[width=0.45\textwidth]{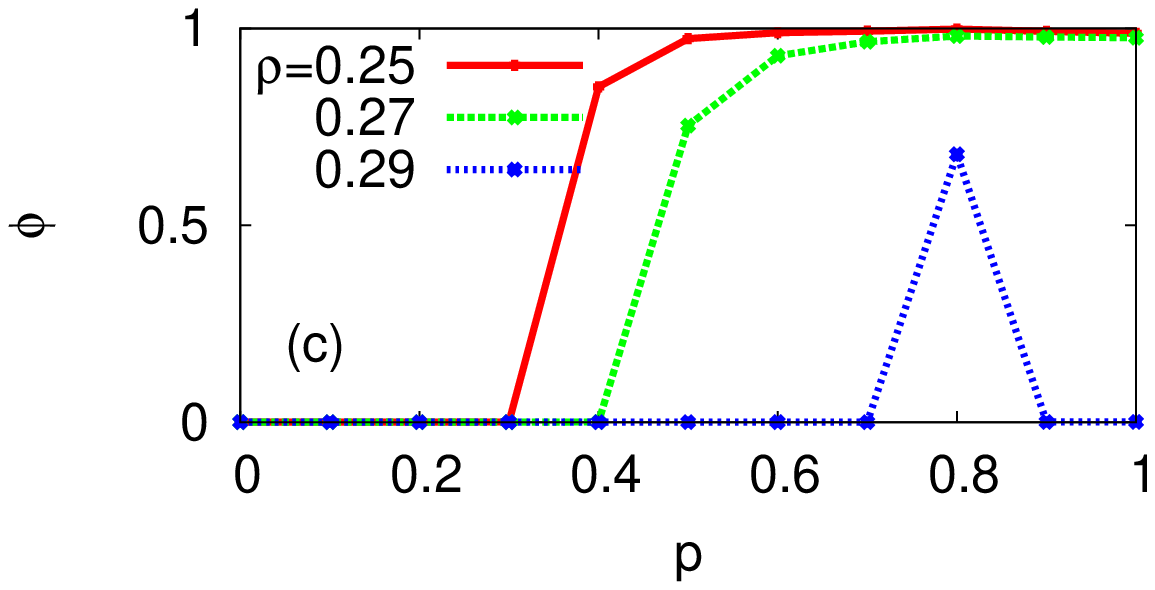}
\includegraphics[width=0.45\textwidth]{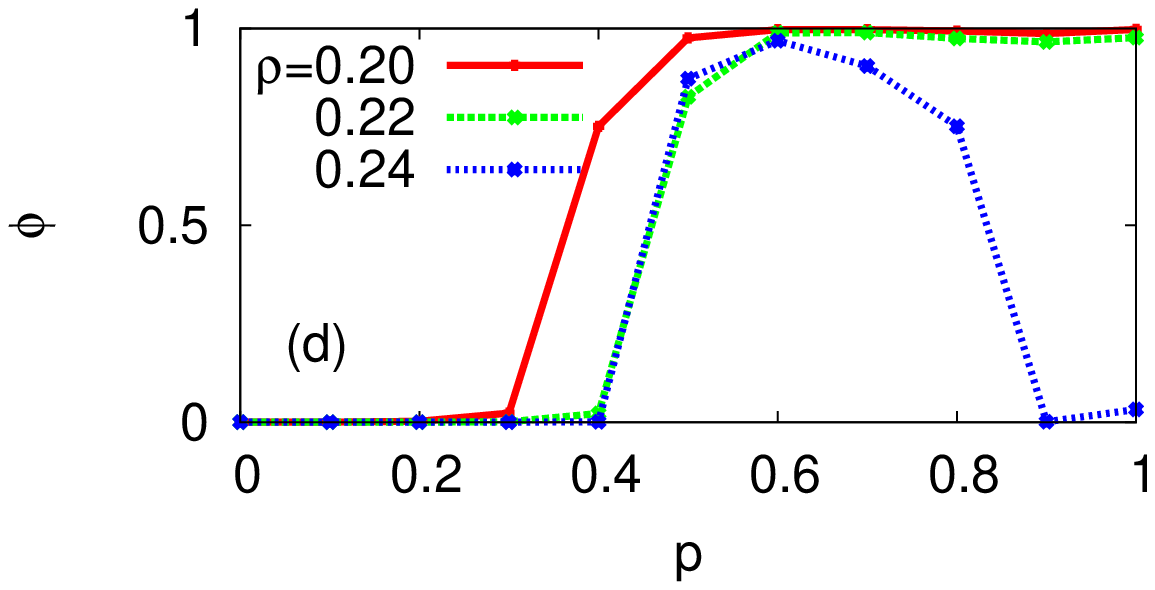}
\caption{(Color online)
Flow $\phi$, as a function of $\rho$ and $p$, with the stopping probability
$s=0$. The road sizes
are (a) $50 \times 200$ and (b) $100 \times 400$, respectively. Each point
is obtained from $10^3$ random initial conditions.
Completely black region indicates that the
system is in a traffic jam, one of
the steady states, while all the agents move freely when the
parameters $p$ and $\rho$ are within
the white region corresponding to $\phi = 1$.
$10^3$ samples.
(c) Sectional plots of $\phi$ with fixing $\rho$ in $50 \times 200$ and (d)
in $100 \times 400$. The nonmonotonic behavior of $\phi$ is more pronounced in
the latter case, which has enough room for developing spatial patterns as
described in the text.
}
\label{fig:est}
\end{figure}

A crucial question, when it comes to traffic flow, is under what conditions
the traffic will jam. This usually happens when the traffic gets too
dense. 
Hence, one may expect that there exists a critical traffic density $\rho =
\rho_c$ beyond which the propensity for jamming becomes high. 
In our traffic model we measure the
traffic flow $\phi$, the fraction of agents advancing in its intrinsic
direction, at a given density of agents $\rho$ and average over a large
number of random initializations. Figure~\ref{fig:tran} shows one
example, together with the average time $\tau$ taken to reach a steady
situation, which is either a jam or a steady-state flow.
Close to $\rho_c$ the time to reach a steady situation becomes so
large that we, for practical reasons, introduce a large time cut-off $t_c$ in
the simulations.
As illustrated in Fig.~\ref{fig:tran}, there is a sharp cross over from a
low- to a high-risk jamming at around a well-defined critical density
$\rho_c$. The flow $\phi$ is averaged over $10^3$ samples in this figure 
and determined by how many samples will settle down to the steady flow.
Intuitively one would expect that $\rho_c$ decreases as we make the road
larger for a fixed density of agents and fixed width of the road; any point
along the road is a potential site where a jamming could start and grow into
a road block across the road which implies that the longer the road the
larger the risk for jamming (see, for comparison, Ref.~\cite{biham}). In
Fig.~\ref{fig:est} it is shown that this is also true in the case where the
width and the length increase simultaneously, preserving the geometrical
shape of the road. Since $\rho_c$ decreases as the size of the road
increases (even when the geometrical shape is preserved), we speculate that
the jamming transition has the large-size limit $\rho_c=0$ in our model.
This also implies that the capacity of a road, measured as the
amount of traffic that a road will transmit on average before the traffic
jams, increases less than linearly with the road size.

A striking feature of Fig.~\ref{fig:est} is that $\rho_c$ does not grow
monotonically with the proportion of traffic-rule abiders $p$.
In other words, when only $60\%$ of agents abide by the rule, for example, the
road of $100 \times 400$ has a higher capacity than when $90\%$ abide by the
rule.

\begin{figure}
\includegraphics[width=0.45\textwidth]{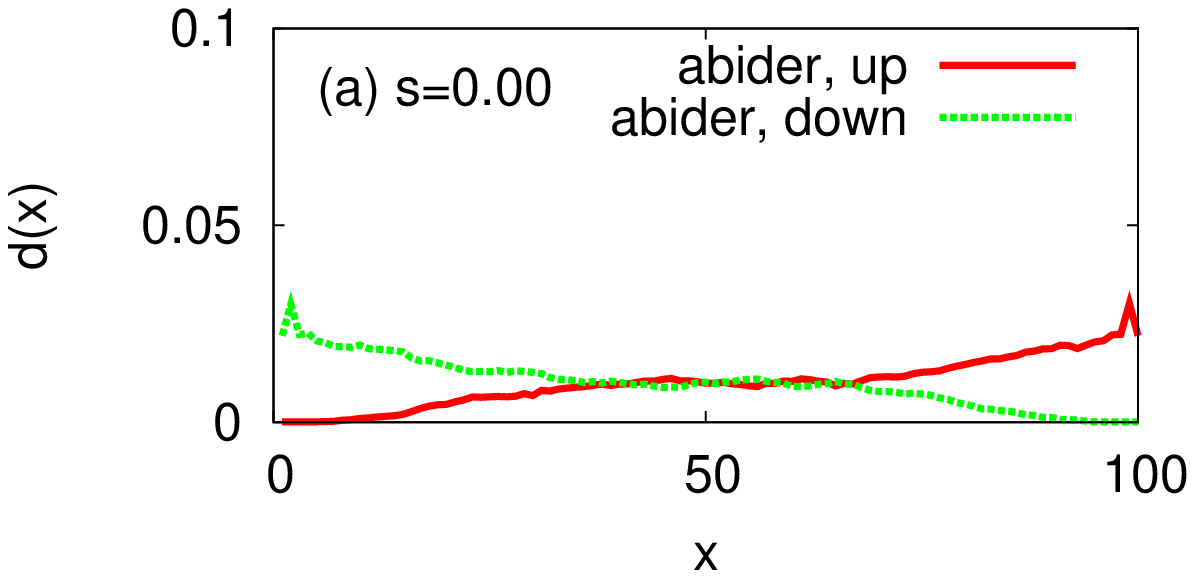}
\includegraphics[width=0.45\textwidth]{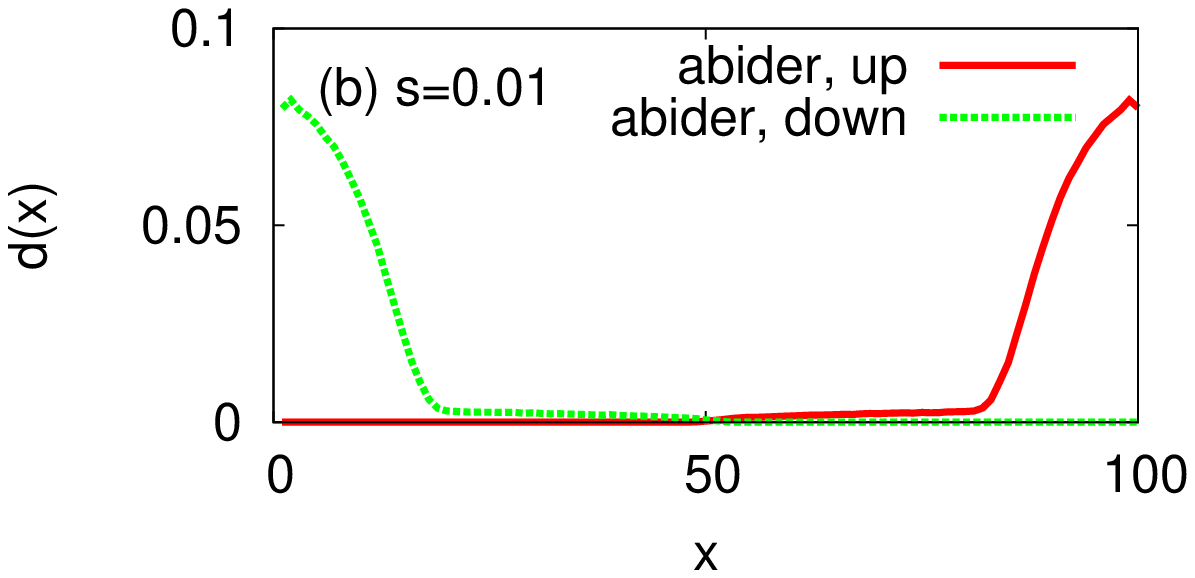}
\caption{(Color online)
Snapshots of density distributions in the long run
when only rule abiders are present ($p=1$).
The average is taken over about $10^3$ samples at $t \ge 6\times 10^4$,
where $X=100$, $Y=400$, and $\rho=0.2$. The stopping probability is
set as (a) $s=0$ and (b) $s=0.01$, respectively.}
\label{fig:b1}
\end{figure}

One possible guess would be that the traffic reduction with abundant
rule abiders is caused by their concentration on the wall sides, since it is
an inefficient use of resources if they are populated only at those parts of
the road. However, that scenario does not explain this phenomenon.
Let us plot the spatial density $d(x)$ for the groups of pedestrians so that
$\sum_x d(x) = 1$ is satisfied for each group.
Rule abiders do not occupy only the sides of the road if the stopping
probability $s$ is zero because then rule abiders have no reason to
move in the lateral direction once forming a lane {\em anywhere} on the road
[Fig.~\ref{fig:b1}(a)].
Even if $s \neq 0$ as in Fig.~\ref{fig:b1}(b),
a high density of agents does not disturb the maximal flow velocity 
due to the recursive update (Sec.~\ref{sub:recur}).

In order to gain some further insight into the mutual effect between
rule abiders and rule ignorers, we have studied the spatial flow structure
in more detail.
To this end it is convenient to include a tiny nonzero stopping probability
$s$ (we use $s=0.01$ in the simulations) 
for the following reason: whenever an agent stops, agents colliding from
behind try to step aside as prescribed by the moving code. Hence a nonzero
stopping probability generates small diffusive processes in the lateral
direction. This helps the system to arrive at a robust
spatial steady state without changing
the numerical results in any essential way.

\begin{figure}
\includegraphics[width=0.45\textwidth]{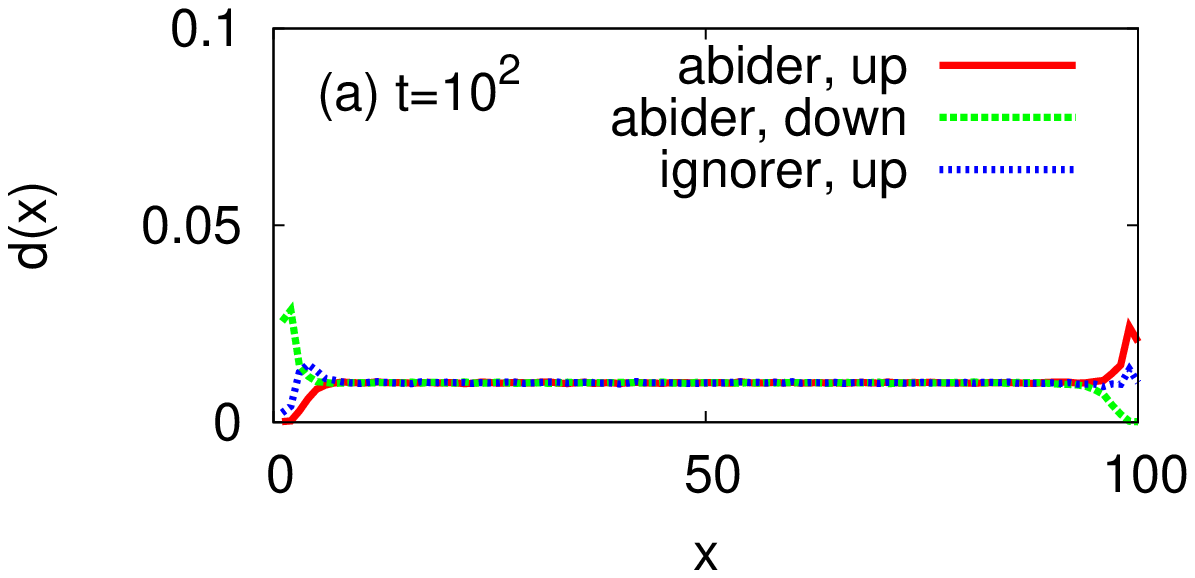}
\includegraphics[width=0.45\textwidth]{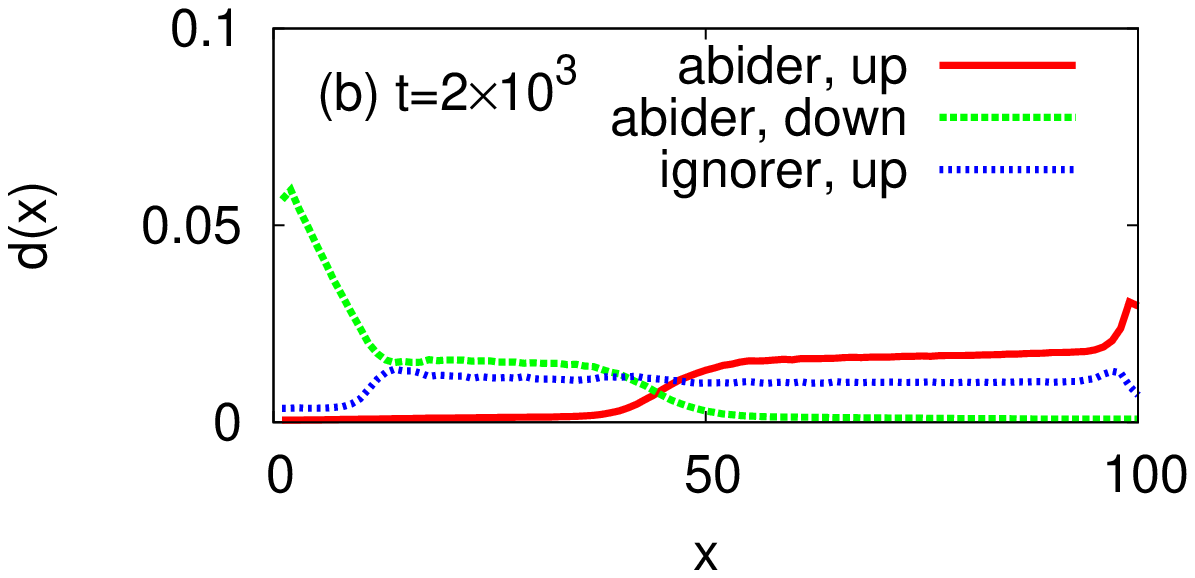}
\includegraphics[width=0.45\textwidth]{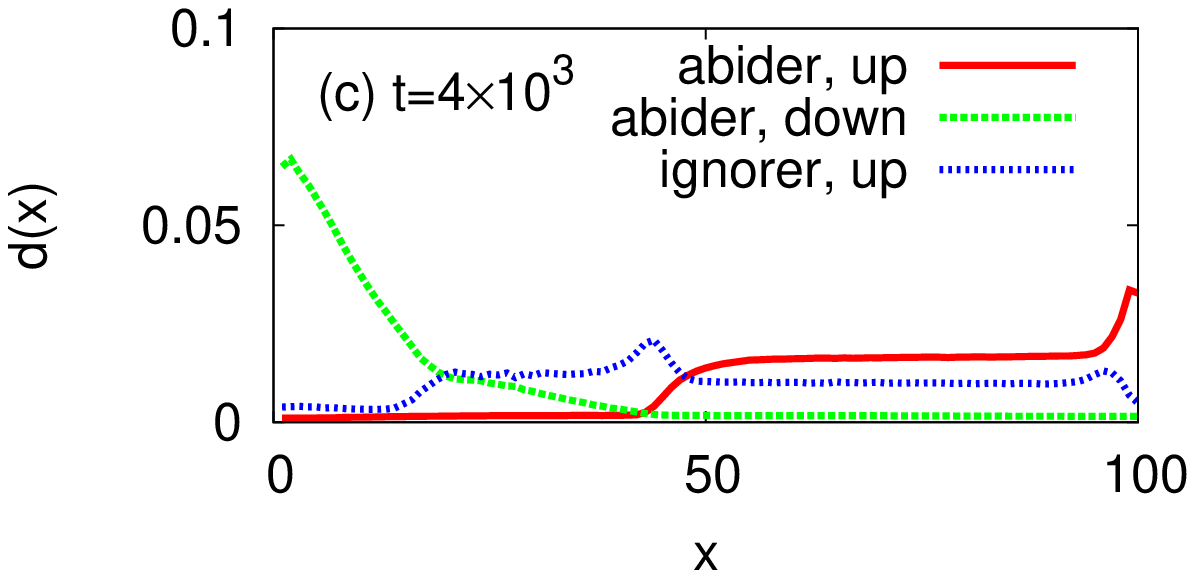}
\includegraphics[width=0.45\textwidth]{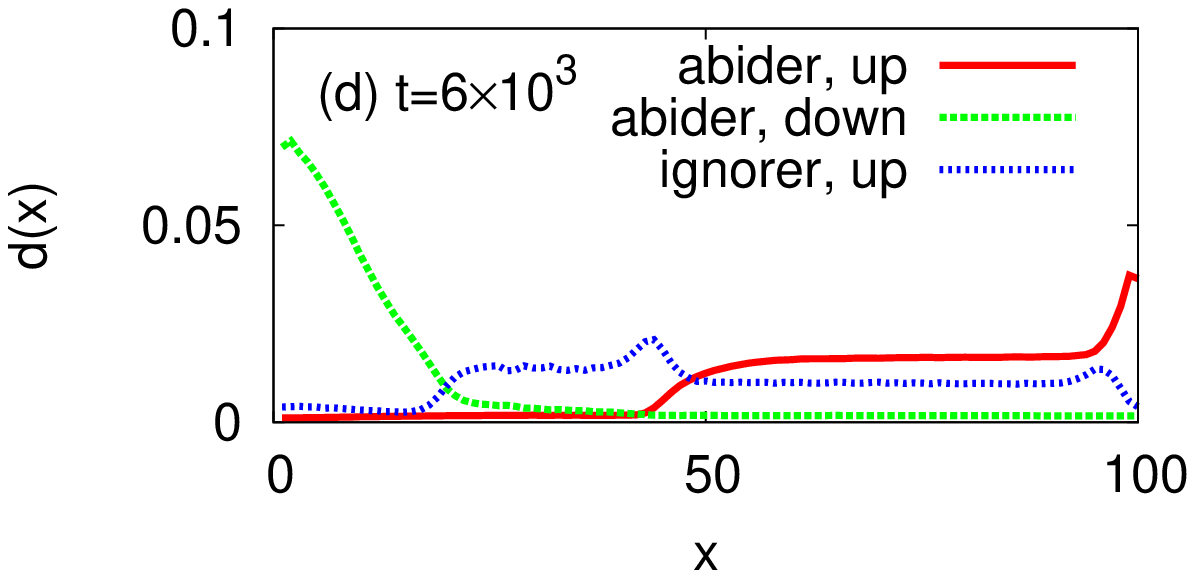}
\caption{(Color online)
Snapshots of density distributions with the condition that
rule ignorers move only upward.
The average is taken over $10^3$ samples,
where $X=100$, $Y=400$, $\rho=0.2$, $p=0.9$, and $s=0.01$.
(a) $t= 10^2$, (b) $t=2 \times 10^3$, (c) $t=4 \times 10^3$,
and (d) $t=6 \times 10^3$.}
\label{fig:snap3}
\end{figure}

For simplicity we choose the case when 
the rule ignorers are restricted to move upward (Fig.~\ref{fig:snap3}).
According to the moving code, agents moving in opposite directions will
have a stronger interaction than agents moving in the same direction.
Since rule abiders always prefer the right-hand side, the most rapid
process is the pushing of the downward rule abiders to the left side
of the plot in order to avoid agents moving upward
[Fig.~\ref{fig:snap3}(b)]. By symmetry the upward rule abiders get pushed
out of the left region preferred by the downward rule abiders.
However, rule ignorers have no preference between left and right, and as a
consequence they
remain for a longer time in the middle region. Their presence
further pushes the downward rule abiders to the left wall.
Note that the upward rule abiders, on the other hand, have little
interaction with rule ignorers since all of them are basically headed for
the same direction. As these upward rule abiders move very slowly in the
$x$ direction, many of the rule ignorers cannot penetrate into the {\em
right} side, i.e., $x>50$, but remain on the {\em wrong} side of the
road with respect to the traffic rule [Fig.~\ref{fig:snap3}(d)].

\begin{figure}
\includegraphics[width=0.45\textwidth]{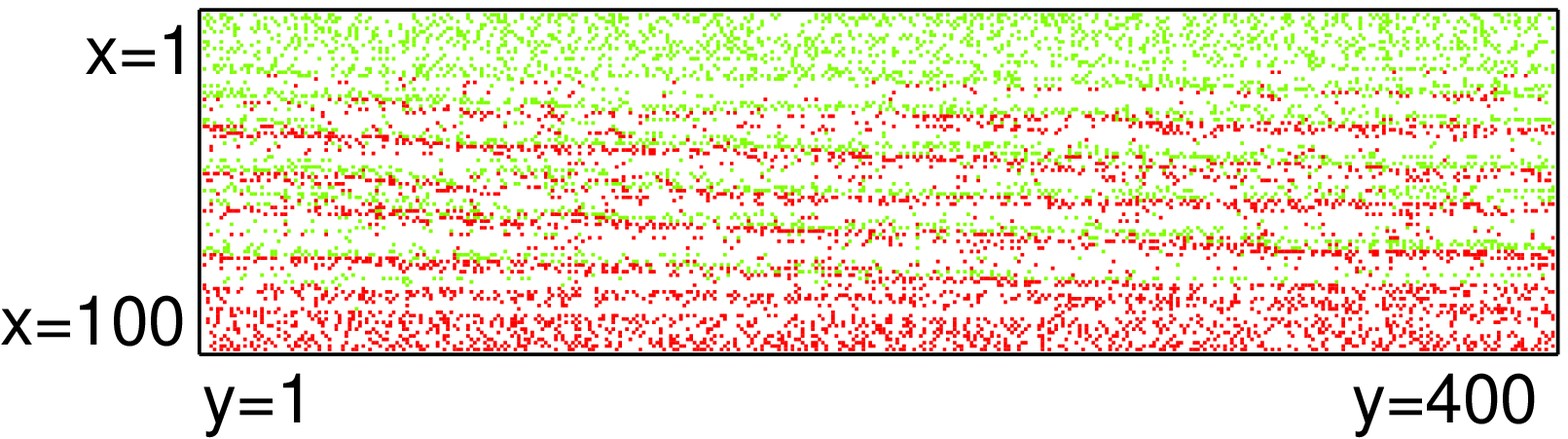}
\caption{(Color online) A typical distribution of agents taken from a sample
at $t=10^3$ with $\rho=0.2$, $p=1.0$, and $s=0.01$, where the road is given
as $100 \times 400$. Two intrinsic directions are marked with different
colors so that red (dark) dots mean upward rule abiders and green (bright)
dots mean downward rule abiders.
}
\label{fig:row}
\end{figure}

\begin{figure}
\includegraphics[width=0.45\textwidth]{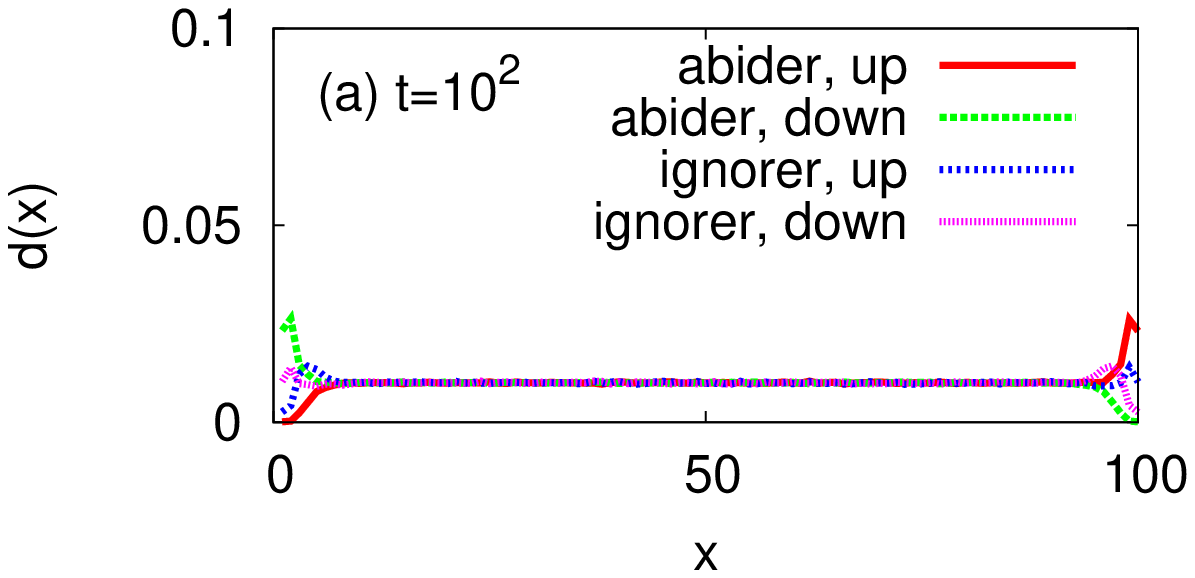}
\includegraphics[width=0.45\textwidth]{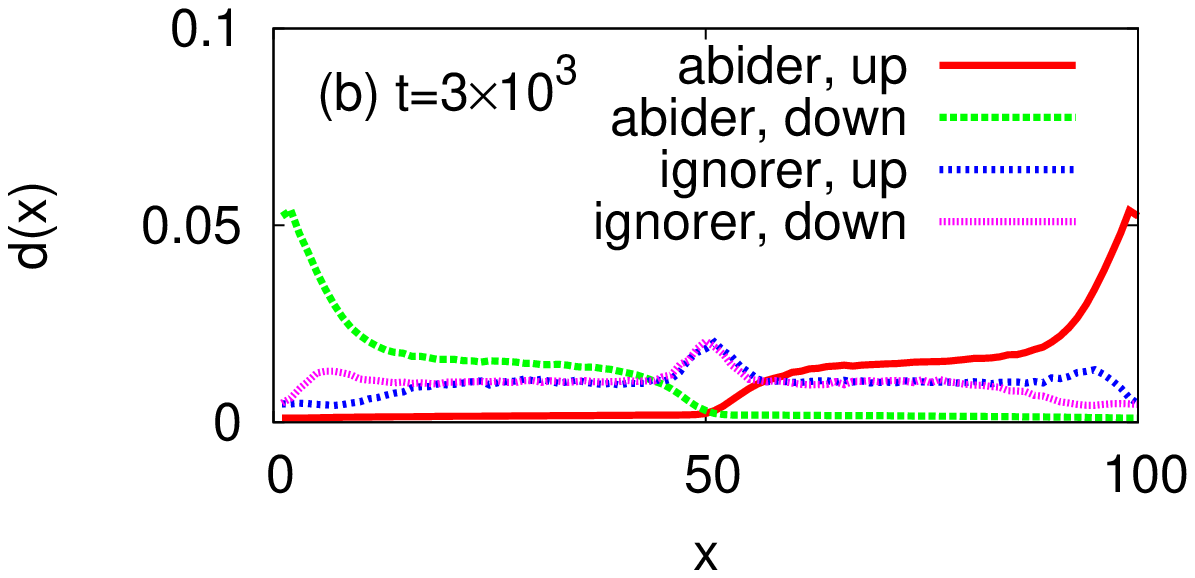}
\includegraphics[width=0.45\textwidth]{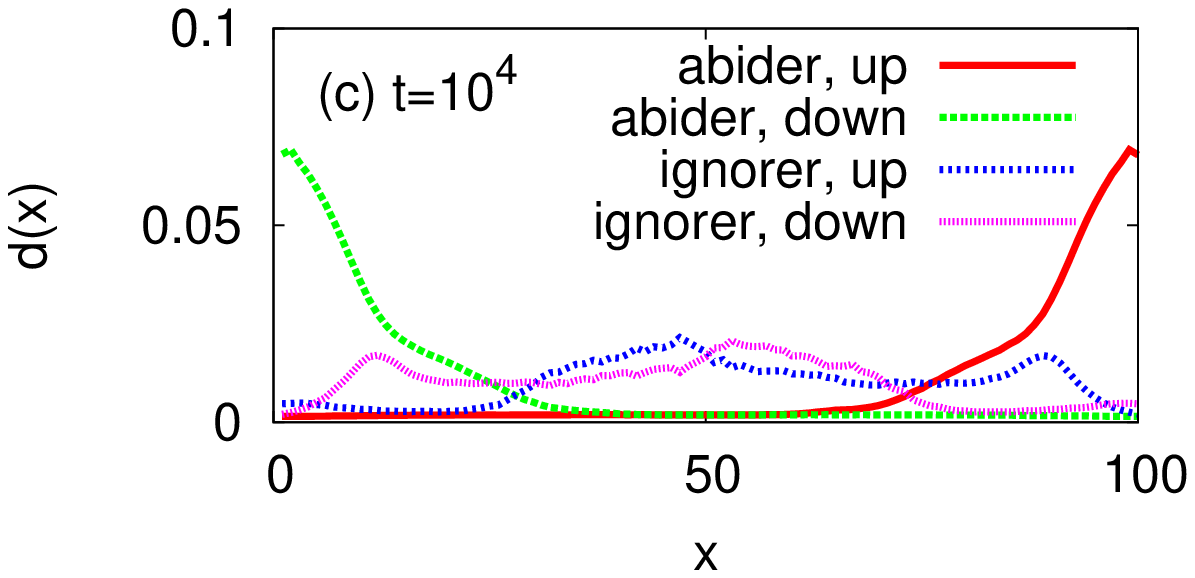}
\includegraphics[width=0.45\textwidth]{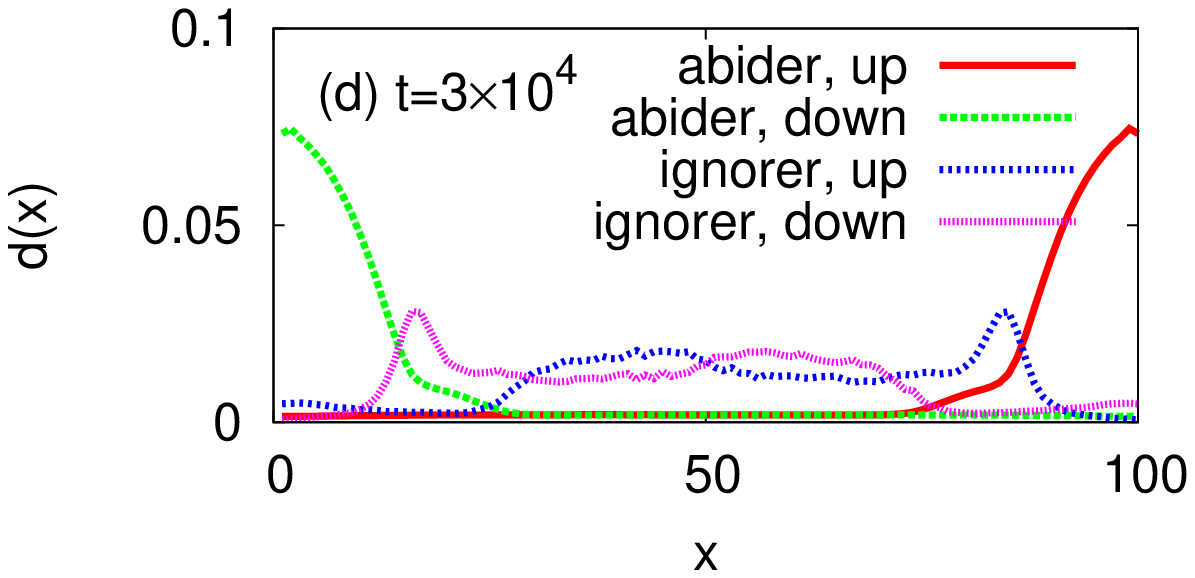}
\caption{(Color online)
Snapshots of density distributions, when both of upward and
downward rule ignorers are present.
The average is taken over $10^3$ samples,
where $X=100$, $Y=400$, $\rho=0.2$, $p=0.9$, and $s=0.01$.
(a) $t=10^2$, (b) $t=3 \times
10^3$, (c) $t=10^4$, and (d) $t=3 \times 10^4$.}
\label{fig:snap4}
\end{figure}

Also in the case which includes downward rule ignorers in addition,
rule abiders are
more quickly evacuated from the central part of the road in the presence of
rule ignorers. In addition, rule ignorers play an important role in
smoothing out uneven agent concentrations on the road. These uneven
concentrations
arise because the upward rule abiders have an average drift toward the
right side of Fig.~\ref{fig:snap3}, when starting from a random initial condition, while
downward
rule abiders drift in the opposite direction. As a consequence they interact
and usually form long narrow trains in the central part of
the road (Fig.~\ref{fig:row}). This leads to high local concentrations on
the road from which jamming can start and develop. However, 
with a sufficient number
of rule ignorers, these trains are broken into a more evenly
distributed concentration, reducing the risk of jamming.

Snapshots of the flow development for the case including downward
rule ignorers as well as upward rule ignorers are shown in
Fig.~\ref{fig:snap4}.
One notable point is that the
rule ignorers tend to make a {\em backflow} against the rule abiders.
One sees
abundance of upward rule ignorers on the left-hand side of the median line
at $x=X/2$ 
and conversely
abundance of downward rule ignorers on the right-hand side of the median
line [Fig.~\ref{fig:snap4}(d)].
The upward (downward) backflow is developed by rule ignorers who are
repelled by downward (upward) movers gathering densely beside the left
(right) wall.
This is a numerically stable pattern
which is rather unexpected from an intuitive point of view.

\begin{figure}
\includegraphics[width=0.45\textwidth]{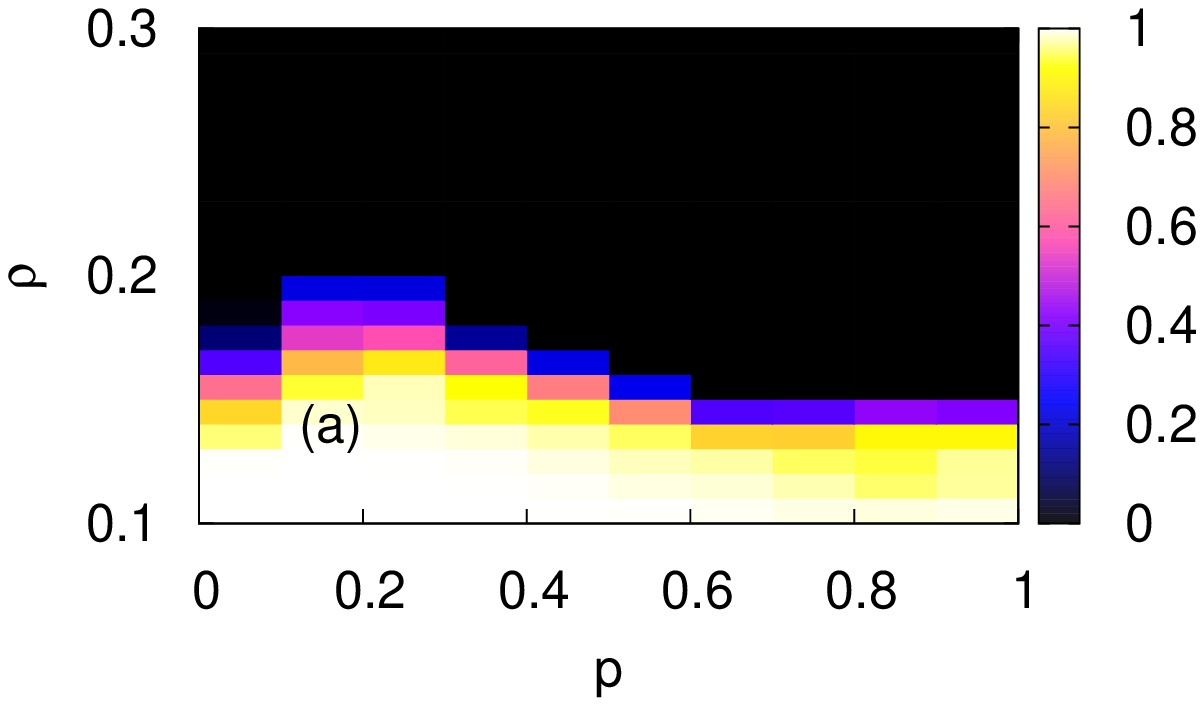}
\includegraphics[width=0.45\textwidth]{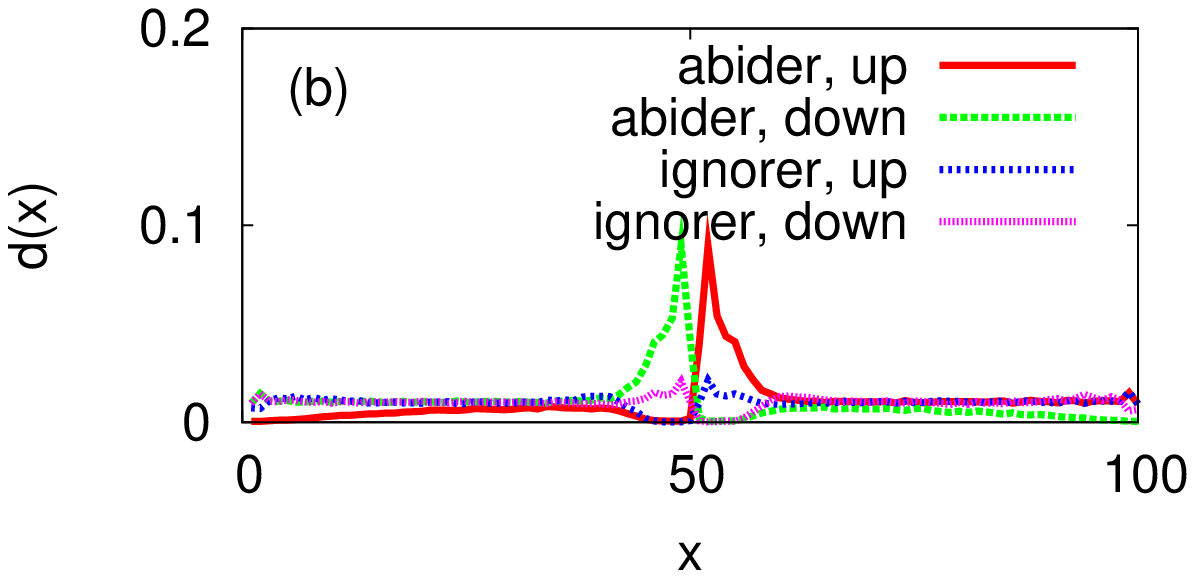}
\caption{(Color online)
(a) Flow $\phi$, as a function of $\rho$ and $p$, when the traffic rule is
applied only to rule abiders on the wrong side of the road.
In accordance to Fig.~\ref{fig:est}, we take averages over $10^3$ samples
with a size of $100 \times 400$ and set $s=0$. (b) A snapshot of the
spatial distribution for $\rho=0.1$ and $p=0.9$, taken at $t=6 \times 10^4$.
This shape maintains itself with time for $s=0$, while for a nonzero $s$,
the peaks on the middle are slowly relaxed outward.
}
\label{fig:r3}
\end{figure}

Finally, we have tested how much the result changes with a variation in the
moving code. The traffic rule becomes softened so that it acts only to
rule abiders on the {\em wrong} side of the road. If standing on the
right side, even rule abiders will be just the same as rule ignorers.
This rule was designed to check whether the excessive concentration on the
wall sides, as in Fig.~\ref{fig:b1}(b), could be relaxed without altering
the qualitative results.
Numerical simulations show that the nonmonotonic behavior of $\rho$ still
remains and the optimal $p$ is even lowered than before
[Fig.~\ref{fig:r3}(a)]. A closer look indicates that jamming is
likely to develop around the median line $x=X/2$, since the momentum of
rule abiders to the right side disappears while crossing the line,
leading to congestion at that point [Fig.~\ref{fig:r3}(b)].

\section{Summary}
\label{sec:dis}

In our numerical simulation based on the CA approach, we have observed that the
jamming transition density on two-dimensional planes does not monotonically
increase with the fraction of rule abiders. It implies that a certain amount
of rule ignorers may diminish the propensity for jamming by diminishing the
risk for high local traffic concentrations.
In contrast to the coordination game, which presumes two rational players
acting to maximize the gain by either abiding to or breaking the
traffic rule, the situation on a large road is generally more complex; it
involves
interactions among agents which lead to nontrivial flow patterns in a long time.
Our result suggests that there are situations when abiding too strictly by a
traffic rule could lead to a jamming disaster which would be avoided if some
people just ignored the traffic rule altogether.

One should note that this is drawn by our model system under certain
conditions, which captures only a part of the pedestrian dynamics. Our
observation demonstrates one possible complexity of the pedestrian
problem that even such simple agents could lead to an unexpected
global behavior. It also supports to some extent a hypothesis that
the most successful behavior in social or biological systems
is achieved when both of the regular and random factors are incorporated,
which should be further examined in future research.

\acknowledgments
S.K.B. and P.M. acknowledge the support from the Swedish Research Council
under Grant No. 621-2002-4135, and B.J.K. is supported by the Korea
Science and Engineering Foundation (KOSEF) through the Research and
Education Program in 2007 and partially under Grant No. R01-2007-000-20084-0.
This research was conducted using the resources of High Performance
Computing Center North (HPC2N).

\end{document}